\documentstyle[11pt]{article}
\catcode`\@=11
\begin{document} \openup6pt

\title{PROBABILITY FOR  PRIMORDIAL BLACK HOLES PAIR IN 1/R GRAVITY}

\author{Dilip Paul\thanks{permanent address : Khoribari H.S. School, Khoribari, Dist. : Darjeeling } and  Bikash Chandra Paul\thanks{ e-mail : bcpaul@nbu.ernet.in }   \\
   Department of Physics, North Bengal University, \\
   Siliguri, Dist. Darjeeling, Pin : 734 013, India }

\date{}

\maketitle

\begin{abstract}

The probability for quantum creation of an inflationary universe with a
 pair of black holes in $1/R$ -  gravitational theory has been studied. 
Considering a gravitational action which includes  a cosmological constant ($\Lambda$) in addition to $ \delta \; R^{- 1} $ term, the probability  
 has been evaluated in a semiclassical approximation with Hartle-Hawking boundary condition.  
We obtain instanton solutions determined by the parameters $\delta$ and $\Lambda$ satisfying the constraint $ \delta \leq \frac{4 \Lambda^{2}}{3}$. However, 
we note  that two different classes of instanton solutions exists in the region $0 < \delta < \frac{4 \Lambda^{2}}{3}$. The probabilities of creation of such configurations are evaluated.
It is found that the probability of  creation of a universe with a pair of black holes is strongly suppressed  with a  positive cosmological constant   except in one case when $0 < \delta < \Lambda^{2}$. It is  also found that gravitational instanton solution is  permitted even with $\Lambda = 0$ but one has to consider $\delta < 0$. However, in the later case   a universe with a pair of black holes is less probable. 
\\

PACS No(s). : 04.70.Dy, 98.80.Bp, 98.80.Cq.

\end{abstract}

\pagebreak

{\bf I. INTRODUCTION : }

It is now generally  believed  from the recent  observations that the present universe is accelerating [1]. It is also believed that our universe might have emerged from an inflationary era in the past. In the last two decades a number of literature  appeared  which explored early universe with an inflationary phase [2]. Inflation not only opens up new avenues in the interface of particle physics and cosmology but also solves some of the outstanding problems not understood in the  standard bigbang cosmology. Most successful  theory  developed so far is based on the dynamics of scalar field with  a suitable potential which behaves like a cosmological constant. In order to explain present acceleration in the universe a number of literature appeared with a new gravitational physics considering theories other than scalar field e.g., Chaplygin gas [3], phantom fields [4] etc.. Moreover, it has been realized that modification of the  Einstein-Hilbert action with higher order terms
in curvature invariants  that are become effective in the  high curvature region  are important in cosmological  model building. The modified theories permit inflation in the early epoch [5]. In the same way it is important to explore modification of the Einstein gravitational action with terms that 
might be  important at extremely low curvature  region  to explain the present cosmic acceleration.  There are attempts to explain the cosmic speed up with a modification of the Einstein Hilbert action. Recently, Carroll {\it et al.} [6] suggested a gravitational action of the form
\begin{equation}
I = \frac{1}{2 \kappa^{2}} \int d^{4} x \sqrt{- g} \left[ R -  \; \frac{\mu^{4}}{R} \right],
\end{equation}
where $R$ is the Ricci Scalar, $\kappa^{2} = 8 \pi G $, and $\mu$ is a mass scale of the order of  Hubble scale. Cosmological models with a modification of the  above action has been used to construct an alternative to dark energy and dark matter models [7]. It is also important to explore the cosmological issues in the framework of new theories which are important at the present epoch. The important astrophysical objects,  for example black holes need  to be investigated in this theory. The mass of these objects may be greater than
 the solar mass or even less. It is known in stellar physics that a 
blackhole is the ultimate corpse of a collapsing star when its mass exceeds 
twice the  mass of the Sun. Another kind of black holes are 
important in cosmology which might have formed due to quantum fluctuations 
of matter distribution in the early universe. These are termed as topological
 blackholes having mass many times smaller than the the solar mass.  In particular these are important 
from the view point of Hawking radiation [8]. Bousso and Hawking [9] 
(in short, BH) calculated the probability of the quantum creation of 
a universe 
with a pair of primordial black holes in  (3 + 1) dimensional universe. 
 In the paper they considered two different 
Euclidean space-time : (1) a universe with space-like sections with 
 $S^{3}$ - topology  and (2) a universe with space-like section with 
$ S^{1} \times S^{2}$- topology, as is obtained in the Schwarzschild- de Sitter
 solution.
The first kind of spatial structure describes an 
inflationary (de Sitter) universe without  black hole while the second kind
 describes a Nariai 
universe [10], an inflationary universe with a pair of black holes.  
BH considered in their paper a theory  with  a massive scalar field which provided an 
effective cosmological constant for a while through a slow-rolling potential 
(mass-term).
 Chao [11] has studied the creation of a primordial black hole and
 Green and Malik [12] studied the primordial black holes production during
 reheating. Paul {\it et al.} [13] following the approach of BH studied the
 probability
of creation of PBH including $R^{2}$-term in the Einstein action and found
that the probability is very much suppressed in the $R^{2}$ -theory. 
Paul and Saha [14] studied probability of creation of a pair of black hole with higher order Lagrangian i.e., considering  higher loop contributions 
into the effective action that are higher than 
quadratic in $R$. In this paper we  investigate a pair of  black holes in the modified gravity given by action (1) in the presence of a cosmological constant.

Consequently it is also important to study the effects of 
these terms to study quantum creation of a universe with a pair of PBH.  
We calculate the probabilities for the creation  of a universe with two
 types of topology namely, $ R \times S^{3} $ -  topology and 
$R \times S^{1} \times S^{2}$ - topology, where the later accommodates
a pair of primordial black holes. 
To calculate the probabilities for these spatial topologies,   we use a 
semiclassical approximation for the evaluation of  the 
Euclidean path integrals. The condition that a classical spacetime 
should emerge, to a good 
approximation, at a large Lorentzian time was selected by a choice of the
 path of the time parameter $\tau$ along the $\tau^{Re}$ axis 
from 0 to $\frac{\pi}{2H}$ and 
 then continues along the $\tau^{im}$ axis. 
The  wave-function of the universe in the semiclassical approximation  is 
given by
\begin{equation}
\Psi_{o} [ h_{ij}  , \Phi_{\partial  M} ]   \approx \sum_{n} A_{n}
 e^{- I_{n}}
\end{equation}
where the sum is over the saddle points of the path integral, 
and $I_{n}$
denotes the corresponding Euclidean action.  The probability measure of the
creation of PBH is
\[
P[ h_{ab} , \Phi_{\partial M} ] \sim  e^{ - 2 \;  I^{Re}}
\]
where $h_{ab}$ is the boundary metric and  $I^{Re}$  is the real part of
the action corresponding to the dominant saddle point, i.e. the classical
solution satisfying the Hartle-Hawking (henceforth, HH) boundary
conditions [15].  It was believed that all inflationary models lead
to $ \Omega_{o} \sim 1$ to a great accuracy. This view was modified after it
was discovered  that there is a special class of inflaton effective 
potentials which
may lead to a nearly homogeneous open universe with  $ \Omega_{o} \leq 1$ 
at the present epoch. Cornish {\it et al.} [16, 17] studied the problem of 
pre-inflationary
homogeneity and outlines the possibility of creation of a small, compact,
 negatively
curved universe. We show that a universe with $S^{3}$-topology may give 
birth to
an open inflation similar to that obtained in {\it Ref.} 18.

The paper is organised as follows : in sec. II  we  write the gravitational
 action for
a higher derivative theories and obtain gravitational instanton solutions 
and
in sec. III we use the action to estimate the relative probability of the 
two types of the universes and in sec. IV we give a brief discussion.

{\bf II. GRAVITATIONAL INSTANTON SOLUTIONS   
WITH OR 
WITHOUT A PAIR OF PRIMORDIAL BLACK HOLES : }

We consider a  Euclidean action   which is
given by
\begin{equation}
I_{E} =  - \frac{1}{16\pi} \int d^{4} x \sqrt{g} \;   f(R)  - 
\frac{1}{8\pi} \int_{\partial M} d^{3} x \sqrt{h}\;  K f'(R) 
\end{equation}
where $g$ is the 4-dimensional Euclidean metric, $f(R) = R +  \frac{\delta}{ R  }  
 - 2 \Lambda $, $R $ is the Ricci scalar, $\delta$ is a dimensional parameter
 and $\Lambda$ is the 4-dimensional cosmological constant.
In the gravitational surface term, $h_{ab}$ is the boundary metric and 
$K = h^{ab} K_{ab}$ is the trace of the second fundamental form of the
 boundary
$\partial M$ in the metric. The second term is the contribution from
 $\tau = 0$ back in the action. It vanishes for a universe with $S^{3}$ 
topology, but gives a non vanishing contribution for 
$S^{1} \times S^{2}$-topology, with the gravitational instanton discussed here.

{\bf (A)  Topology $S^{3}$, the de Sitter spacetime :}

In this section we study vacuum solution of the modified  Euclidean Einstein action with a cosmological constant in four dimensions. We now look for a solution 
with spacelike section having $S^{3}$ - topology and the corresponding 
four dimensional metric ansatz which is given by
\begin{equation}
 \;
ds^{2} = d\tau^{2} + a^{2} (\tau) \left[ dx_{1}^{2} + sin^{2} x_{1} \;
d\Omega_{2}^{2} \right]
\end{equation}
where $a(\tau)$ is the scale factor of a four dimensional universe and
 $ d\Omega_{2}^{2} $ is a line element on the unit two sphere.
The scalar curvature is given by
\[
R = -  6 \left[ \frac{\ddot{a}}{a} + \left(
    \frac{\dot{a}^{2}}{a^{2}} - \frac{1}{a^{2}} \right) \right] .
\]
where an overdot denotes differentiation with respect to $\tau$. We rewrite 
the action (3), including the constraint through a Lagrangian multiplier 
$\beta$ and obtain 
\begin{equation}
I_{E} = -  \frac{\pi}{8}  \int 
         \left[ f(R) a^{3} - \beta \left( R + 6 \frac{\ddot{a}}{a} + 
6 \frac{\dot{a}^{2} -1}{a^{2}} \right) \right] d \tau 
- \frac{1}{8 \pi} \int_{\partial M} d^{3} x \sqrt{h} K f'(R).      
\end{equation}
Varying the action w.r.t. R, we determine
\begin{equation}
\beta = a^{3} f'(R).
\end{equation}
Substituting the above equation, treating $a$ and $R$ as independent 
variables we get 
\[
I_{E} = -  \frac{\pi}{8}  \int_{\tau = 0}^{\tau_{\frac{\pi}{2 H}} } 
         \left[ a^{3} f(R) - f'(R) \left( a^{3} R - 6  a \dot{a}^{2} - 6 a 
\right)  + 6 a^{2} \dot{a} \dot{R} f''(R)  \right] d \tau
\]
\begin{equation}
-  \frac{3 \pi}{4} \left[  \dot{a} a^{2} f'(R) \right]_{\tau = 0},
\end{equation}
here we have eliminated $\ddot{a}$ term in the action  by integration by parts. The field 
equations are  now obtained by varying the action with respect to $a$  and
 $R$ respectively, giving
\begin{equation}
f''(R) \left[ R + 6  \frac{ \ddot{a}}{a}    + 
6 \frac{\dot{a}^{2} - 1}{a^{2}} \right]  = 0 ,
\end{equation}
\begin{equation}
2 f'''(R) \dot{R}^{2} + 2 f''(R) \left[ \ddot{R} + 2 \frac{\dot{a}}{a} 
\dot{R} \right] + f'(R) \left[ 4 \frac{ \ddot{a}}{a} + 2 
\frac{ \dot{a}^{2}}{a^{2}} - \frac{2}{a^{2}} + R \right] - f(R)  = 0.
\end{equation}
We now consider $f(R) = R + \frac{\delta}{R}  - 2 \Lambda $ and
 obtain  an instanton solution which is given by
\begin{equation}
a = \frac{1}{H} \; \sin  H \tau 
\end{equation}
where
$ R = 12 H^{2} $ and $H$ is determined from the constraint equation 
\begin{equation}
48  H^{4} - 16 \Lambda H^{2} + \delta = 0. 
\end{equation}
We note that (i) two unequal values of $H$ are permitted when $0 < \delta < \frac{4 \Lambda^{2}}{3}$, (ii) $H
^{2} = \frac{\Lambda}{3}$ for $\delta = 0$, and (iv)$\Lambda = 0$ permits a real value for $H^{2} $ when  $\delta < 0$.
It is evident that the instanton solution obtained here satisfies the HH no boundary conditions viz., 
$a(0) = 0 $, $ \dot{a} (0) = 1$. One can choose a path along the $\tau^{Re}$
axis to  $\tau = \frac{\pi}{2 H}$, the solution describes half of the 
Euclidean
de Sitter instanton $S^{3}$.  Analytic continuation of the metric (3) to 
Lorentzian region, $x_{1} \rightarrow \frac{\pi}{2} + i \sigma $, gives
\begin{equation}
ds^{2} = d \tau^{2} + a^{2}(\tau) \left[ - d\sigma^{2} + \cosh^{2} \sigma
\;  d \Omega_{2}^{2} \right]
\end{equation}
which is a spatially inhomogeneous de Sitter like metric. However, if one 
sets $\tau = i t $ and  $ \sigma = i \frac{\pi}{2} + \chi $, the metric 
becomes
\[
ds^{2} = -  dt^{2} + b^{2}(t) \;  [ d\chi^{2} + \sinh^{2} \chi
 \;
 d \Omega_{2}^{2} ]
\]
where $b(t) = - \; i \;  a( it )$. The line element (12) now describes  an open 
inflationary  universe.
The real part of the Euclidean action 
corresponding to the solution calculated by following the complex contour
of $\tau $ suggested by BH is given by
\begin{equation}
I^{Re}_{S^{3}} = - \frac{2 \pi}{3  H^{2}} \left[ \frac{ 12 \Lambda H^{2} -  \delta }{ 16 \Lambda H^{2} - \delta } \right].
\end{equation}
With the chosen path for $\tau $,  the solution describes 
half the de Sitter instanton  with $S^{4}$ 
topology, joined to a real Lorentzian hyperboloid of
topology  $R^{1} \times S^{3}$. It can be joined to any boundary satisfying 
the condition 
$a_{\partial M} > 0$ . 
For $a_{\partial M} > H^{- 1}$ , the wave function oscillates and 
predicts a classical
space-time. 
We note the following cases :

$\bullet$  $\delta = 0$, the action reduces to that obtained by BH.

$\bullet$  $0 < \delta < \frac{4 \Lambda^{2}}{3}$,  a realistic solution corresponds to $H_{o}^{2} > \frac{\delta}{12 \Lambda} $.

$\bullet$   $\Lambda = 0$, the action is decided by $\delta$ which is $I = - \frac{8 \pi}{\sqrt{- 3 \; \delta}}$, it is real for $\delta < 0$.

{\bf (B) Topology $ S^{1} \times S^{2} $ :}

In this section we consider  Euclidean Einstein equation
 and look for a universe with $S^{1} \times
S^{2}$ -spacelike sections as this topology accommodates a pair of black 
holes. The corresponding ansatz for (1 + 1 + 2) dimensions is given by
\begin{equation}
ds^{2} = d \tau^{2} + a^{2}(\tau) \; dx^{2} + b^{2}(\tau) \;  d \Omega_{2}^{2} 
\end{equation}
where $a( \tau ) $ is the scale factor of $S^{1}$-surface and $b( \tau )$ is the
scale factor of the two sphere. The metric for the two-sphere is  given by the metric 
\[
d\Omega_{2}^{2} = d \theta^{2} + sin^{2} \theta \; d \phi^{2} 
\]
The scalar curvature is given by
\begin{equation}
R = - \left[ 2 \frac{\ddot{a}}{a} +  4 \frac{\ddot{b }}{b}
     + 2 \left( \frac{\dot{b}^{2}}{b^{2}} - 
   \frac{1}{b^{2}} \right) + 4 \frac{\dot{a} \dot{b}}{a b} \right] .
\end{equation}
The Euclidean action (3) becomes
\[
I_{E} = - \frac{\pi}{2} \int
 \left[  f(R) a b^{2} - \beta \left(R + 2 \frac{\ddot{a}}{a} + 4
\frac{\ddot{b}}{b} + 4 \frac{\dot{a} \dot{b}}{ab} + 2 
\frac{\dot{b}^{2}}{b^{2}} - \frac{2}{b^{2}}\right) \right] d\tau
\]
\begin{equation}
 - \frac{1}{8 \pi} \int_{\partial M} \sqrt{h} \;  d^{3}x \;  K f'(R).
\end{equation}
One can determine $\beta$ as is done before and obtains 
\[
I_{S^{1} \times S^{2}} = - \frac{\pi}{2} 
\int_{\tau = 0}^{\tau_{\partial M}} 
 \left[  f(R) - f'(R) \left( R - 4  \frac{\dot{a} \dot{b}}{ab} - 
2 \frac{\dot{b}^{2}}{b^{2}} - \frac{2}{b^{2}}\right) + 2 f''(R)  
\dot{R} \left( \frac{\dot{a}}{a} + 2 \frac{\dot{b}}{b} \right) \right] 
\]
\begin{equation}
a b^{2} d \tau - \pi \left[ \left( \dot{a} b^{2}  + 2 a b \dot{b} \right) f'(R) 
\right]_{\tau = 0}.
\end{equation}
Variation of the action with respect to $R $, $a$ and $b$
 respectively are given by
the following  equations
\begin{equation}
f''(R) \left[  R + 2 \frac{\ddot{a}}{a}
  + 4 \frac{\ddot{b}}{b} + 4 \frac{\dot{a} \dot{b}}{ab}  
+ 2 \frac{\dot{b}^{2}}{b^{2}} - \frac{2}{b^{2}} \right] = 0,
\end{equation}
\begin{equation}
2  f'''(R) \dot{R}^{2} + 2 f''(R) \left[ \ddot{R} + 2 \dot{R} 
\frac{\dot{b}}{b}
\right] + f'(R) \left[ R + 4 
\frac{\ddot{b}}{b}  +  
2 \frac{ \dot{b}^{2} - 1}{b^{2}} \right] - f(R)  = 0,
\end{equation}
\[
2 f'''(R) \dot{R}^{2} + 2 f''(R) \left[ \dot{R} \left(\frac{\dot{a}}{a} +
\frac{\dot{b}}{b}\right)  + \ddot{R} \right] + f'(R)  
\left[  R + 2  \frac{\ddot{a}}{a} +
 2 \frac{\ddot{b}}{b}  + 2 \frac{\dot{a} \dot{b}}{ab} \right] 
\]
\begin{equation}
- \; f(R)   = 0.
\end{equation}
Let us now consider $f(R) = R + \frac{\delta}{R}  - 2 \Lambda $. Equations (18)-(20) admit an instanton solution  which is given by
\[
a = \frac{1}{H_{o}} \; sin ( H_{o} \tau ) , \; \; \;
b =  H_{o}^{- 1},
\]
\begin{equation}
R = 4 H_{o}^{2}, 
\end{equation}
where $H_{o}$ satisfies the constraint equation 
\begin{equation}
16  H_{o}^{4} - 16 \Lambda  H_{o}^{2} + 3 \delta  = 0.
\end{equation}
 We note the following : (i)   $\delta = 0$, one gets   $H_{o}^{2} = \Lambda$, (ii)  $ 0  < \delta < \frac{4 \Lambda^{2}}{3}$, one obtains   two unequal real values of     $H_{o}^{2}$ are permitted, (iii) $\delta =  \frac{4}{3} \Lambda^{2}$ leads to  $H_{o}^{2} = \frac{\Lambda}{2}$  and (iv) $\Lambda  = 0$, gives  $H_{o}^{2} = \sqrt{ - \; \frac{3 \delta}{16} }$ which is real if   $\delta < 0$.
The above instanton solution satisfies the HH boundary conditions
$a(0) = 0 $, $ \dot{a} (0) = 1,  b(0) = b_{o} $, $ \dot{b} (0) = 0$.
Analytic continuation of the metric (14) to Lorentzian region, i.e.,
$\tau \rightarrow it $ and $ x \rightarrow \frac{\pi}{2} + i \sigma $ yields
\begin{equation}
ds^{2} = -  dt^{2} + c^{2}(t) \; d\sigma^{2} + H_{o}^{-2} \; d \Omega_{2}^{2},
\end{equation}
where $c(t) = - \;  i \;  a( i t )$. In this case the analytic continuation of time
and space
do not give an open inflationary universe. The corresponding Lorentzian 
solution
is given by
\[
a(\tau^{Im}) |_{\tau^{Re} \;  = \frac{\pi}{2H_{o}}} = H_{o}^{-1} \cosh H_{o} \tau^{Im} ,
\]
\[
b(\tau^{Im}) |_{\tau^{Re} \;  = \frac{\pi}{2H_{o}}} = H_{o}^{-1} 
\]
Its space like sections can be visualised as  three spheres of radius $H_{o}^{-1}$
with a {\it hole} of radius $b = H_{o}^{-1}$ punched through the north and south poles. 
The
physical interpretation of the solution is that of two - spheres containing
two black holes at opposite ends. The black holes have the radius $H_{o}^{-1}$ which
accelerates away from each other with the expansion of the universe.
The
real part of the action can now be determined following the contour
 suggested  by BH [2], and it is given by
\begin{equation}
I^{Re}_{S^{1} \times S^{2}} = - \frac{4 \pi}{  H_{o}^{2}} \left[ \frac{4 \Lambda
  H_{o}^{2} - \delta}{16 \Lambda H_{o}^{2} - 3 \delta} \right].
\end{equation}
The solution (24) describes a universe with two black holes at the poles
of a two sphere. It may be pointed out  here that the contribution of the integrand in (17) for the instanton vanishes and the non-zero contribution of the action here arises from the boundary term only. We note the following :

$\bullet $ $\delta = 0$, the action reduces to that obtained by BH.

$\bullet$   $\Lambda = 0$, the action is decided by $\delta$ which is $I = - \frac{16 \pi}{3 \sqrt{- 3 \; \delta}}$, it is real for $\delta < 0$.

$\bullet$  $\delta = \frac{4 \Lambda^{2}}{3}$, the action becomes  $I = - \frac{4 \pi}{3\Lambda}$.

$\bullet$  $\delta < \frac{4 \Lambda^{2}}{3} $,  either $ H_{o}^{2} > \frac{\delta}{4 \Lambda}$ or  $ H_{o}^{2} < \frac{3 \delta}{16 \Lambda}$ allows a negative definite action corresponding to the instanton solution.

{\bf III. PRIMORDIAL BLACK HOLES PAIR CREATION  PROBABILITY :}

In the previous section we have calculated the actions for inflationary 
universe with or without a pair of black holes. We now compare the
 probability
measure  in the two cases.
The probability for creation of a  de Sitter universe 
is determined from the action (7).  The probability for
nucleation of an inflationary universe without  a pair of Black holes is given by
\begin{equation}
P_{S^{3}} \sim \; e^{ \frac{\pi}{36 H^{6}} \left[ 12 \Lambda  H^{2} - \delta \right] }.
\end{equation}
However for an inflationary universe with a pair of black holes the 
corresponding probability of nucleation can be obtained from the
action (17).  The corresponding probability is
\begin{equation}
P_{S^{1} \times S^{2}} \sim e^{ \frac{ \pi}{2 H_{o}^{6}} \left[
 4 \Lambda  H_{o}^{2} - \delta \right] }
\end{equation}
We now describe  spacial cases  :

$\bullet$  $\delta = 0$,  one  recovers the result obtained by 
Bousso and Hawking [9]
\begin{equation}
P_{S^{3}} \sim e^{\frac{3 \pi}{\Lambda}} , \; \; \; P_{S^{1} 
\times S^{2}} \sim e^{\frac{2 \pi}{\Lambda}}.
\end{equation}
Thus with a positive cosmological constant the probability for a
 universe with PBH is less than that without PBH.

$\bullet$ $\delta \neq  0$, since eqs. (11) and (22) are quadratic in $H^{2}$ and $H_{o}^{2}$, there are two branches of solutions admitting two different instanton solutions. In one case  
the probabilities  are given by
\begin{equation}
P_{S^{3}} \sim e^{\frac{6 \pi}{ \left( \Lambda +  \sqrt{ \Lambda^{2} - \frac{3 \delta}{4}} \right)^{3}} \left[  2  \Lambda  \left(  \Lambda  + \sqrt{ \Lambda^{2} - \frac{3 \delta}{4}} \right)  - \delta \right]}
,  \; \; \; 
P_{S^{1} \times S^{2}} 
\sim e^{\frac{4 \pi}{ \left( \Lambda +  \sqrt{ \Lambda^{2} - \frac{3 \delta}{4}} \right)^{3}} \left[  2  \Lambda  \left(  \Lambda  + \sqrt{ \Lambda^{2} - \frac{3 \delta}{4}} \right)  - \delta \right]}
\end{equation}
for $\delta < \frac{4 \Lambda^{2}}{3}$, and in the other case the probabilities are 
\begin{equation}
P_{S^{3}} \sim e^{\frac{6 \pi}{ \left( \Lambda -  \sqrt{ \Lambda^{2} - \frac{3 \delta}{4}} \right)^{3}} \left[  2  \Lambda  \left(  \Lambda  - \sqrt{ \Lambda^{2} - \frac{3 \delta}{4}} \right)  - \delta \right]}
,  \; \; \; 
P_{S^{1} \times S^{2}} 
\sim e^{\frac{4 \pi}{ \left( \Lambda -  \sqrt{ \Lambda^{2} - \frac{3 \delta}{4}} \right)^{3}} \left[  2  \Lambda  \left(  \Lambda  - \sqrt{ \Lambda^{2} - \frac{3 \delta}{4}} \right)  - \delta \right]}
\end{equation}
when $0 < \delta < \frac{4 \Lambda^{2}}{3}$.

It may be pointed out here that the values of $H^{2}$ and $H_{o}^{2}$ corresponding to the probability measure given by eq.(28) are  higher than that in eq.(29).
It is evident from eqs. (28) and (29) that the creation of  a universe without PBH is more probable than with PBH
for  $\delta < \frac{4 \Lambda^{2}}{3} \; $ with a positive cosmological constant. It admits negative $\delta$ also. However,  the probability of creation of a universe with a pair of PBH is favoured in the  first case when one considers a negative cosmological constant. In the second case, the probability of creation of a universe with a pair of PBH is favoured with a positive cosmological constant for $0 < \delta < \Lambda^{2}$, but it is suppressed  for 
$ \Lambda^{2} < \delta < \frac{4 \Lambda^{2}}{3}$. In the second case no instanton solution exists with a negative cosmological constant.
  
 $\bullet$ $\delta = \frac{4 \Lambda^{2}}{3}$, the probabilities are given by 
\begin{equation}
P_{S^{3}} \; \sim  \; e^{\frac{4 \pi}{\Lambda}},  \; \; \; \;  P_{S^{1} \times  S^{2}} \; \sim  \;  e^{\frac{8 \pi}{ 3 \Lambda}}.
\end{equation}
In the above only one kind of instanton solution is permitted with $H = \sqrt{ \frac{\Lambda}{6}}$ in the $S^{3}$-topology and $H_{o} = \sqrt{ \frac{\Lambda}{2}}$ in the $S^{1} \times S^{2}$ -topology.

$\bullet$  $\Lambda = 0$, the probabilities are given by 
\begin{equation}
P_{S^{3}} \; \sim  \; e^{\frac{16 \pi}{\sqrt{- 3 \delta}}},  \; \;  \; \; P_{S^{1} \times S^{2}} \; \sim  \; e^{\frac{32 \pi}{ 3 \sqrt{- 3 \delta}}}.
\end{equation}
in this  case only negative values are permitted  for the  parameter $\delta$ in the action which  reduces to the action that considered by Carrol {\it et al.} [6].  

{\bf IV. DISCUSSIONS : }

In this paper  we have evaluated the probability for primordial
black holes pair creation in a modified theory of gravity. In
section II, we have obtained the gravitational instanton solutions in  the two 
cases : (i)  a universe with $R \times S^{3}$ - topology  and  (ii) a universe with  $ R \times S^{1} \times S^{2} $ - topology respectively. The 
Euclidean  action is then evaluated corresponding to the instanton solutions. For a non zero $\delta$ in the gravitational action, we obtain two classes of instanton solutions differing in the values of  the radius of the sphere $H^{ - 1}$ and $H_{o}^{- 1}$ respectively.
 We found  that the probability of a
 universe
with $R \times S^{3} $ topology turns out to be much lower than a universe
 with topology $ R \times S^{1} \times S^{2} $ 
in the modified theory unless 
$0 < \delta < \frac{4 \Lambda^{2}}{3}$ with a positive cosmological constant in the case of lower instantonic radius ($H^{- 1}, H_{o}^{- 1}$)   and   in the case of higher instantonic radius ($H^{- 1}, H_{o}^{- 1}$) with $ \Lambda^{2} < \delta < \frac{4 \Lambda^{2}}{3} $ with a positive cosmological constant. It may be mentioned here that one gets a regular instanton in 
$S^{4} $ - topology
if there are no black holes. The existence of black holes  
restricts such a regular topology. 
The results obtained here on the probability of creation of a universe with a  pair of primordial black holes are found to be strongly 
suppressed depending on the parameter $\delta$ determined by the cosmological constant in some cases which are presented here. We note an interesting solution here in the framework of the modified gravitational action with inverse power of $R$ -theory, which admits  de Sitter instantons with $S^{3}$ and $S^{1} \times S^{2}$ topologies even without a cosmological constant. In this case the action is similar to that considered by Carroll {\it et al.} [6].
It is interesting to note here that analytic continuation of a 
$R \times S^{3}$ metric considered here
to Lorentzian region leads to an open 3 - space.  One
obtains Hawking-Turok [18] type open inflationary universe in this case.
In the 
other type of topology  an open inflation section of the universe  is not permitted.
A detail study of an open inflationary universe will be discussed elsewhere.
Thus in a modified Lagrangian with inverse power in  $R$-theory, quantum creation of PBH seems 
to be suppressed in the minisuperspace cosmology for some values of the 
parameters in the action, which are determined here.
Another new result obtained here is that gravitational instanton solution may be obtained even with a negative cosmological constant which is not permitted in the case considered by BH [9].

\vspace{0.5 in}

{\bf {\it Acknowledgement  :}}
BCP gratefully acknowledges valuable discussions with Naresh Dadhich. DP  would like to thank S. Mukherjee for providing facilities of the IUCAA Reference Centre at North Bengal University to carry out a part of the work. This  work is partially supported by University Grants Commission, New Delhi.
\pagebreak

\end{document}